\documentclass[aps,prb]{revtex4}
\usepackage{amsmath,amssymb}
\usepackage{graphicx}
\usepackage{dcolumn}
\usepackage{bm}
\usepackage{color}
\newcommand{\mbb}{\mathbb}
\newcommand{\mc}{\mathcal}

\newcommand{\tet}{\texttt}
\newcommand{\pr}{\partial}

\begin{document}
\title{
Adjustable propagating plasmons in $\alpha - \mc{T}_3$ lattice-based armchair nanoribbons
}

\author{Andrii Iurov$^{1}\footnote{E-mail contact: aiurov@unm.edu, theorist.physics@gmail.com}$,
Liubov Zhemchuzhna$^{1,2}$,
Paula Fekete$^{3}$, 
Godfrey Gumbs$^{2,4}$, 
Danhong Huang$^{5,6}$
Farhana Anwar$^{7}$,
Dipendra Dahal$^{8}$, 
and 
Nicholas Weekes$^{1,9}$, 
}
\affiliation{
$^{1}$Department of Physics and Computer Science, Medgar Evers College of City University of New York, Brooklyn, New York 11225, USA\\
$^{2}$Department of Physics and Astronomy, Hunter College of the City University of New York, 695 Park Avenue, New York, New York 10065, USA\\
$^{3}$US Military Academy at West Point, 606 Thayer Road, West Point, New York 10996, USA\\ 
$^{4}$Donostia International Physics Center (DIPC), P de Manuel Lardizabal, 4, 20018 San Sebastian, Basque Country,
Spain\\
$^{5}$Air Force Research Laboratory, Space Vehicles Directorate, Kirtland Air Force Base, NM 87117, USA\\
$^{6}$Center for High Technology Materials, University of New Mexico, 1313 Goddard SE, Albuquerque, NM, 87106, USA\\
$^{7}$Lawrence Berkeley National Laboratory, 1 Cyclotron Rd, Berkeley, CA 94720, USA\\
$^{8}$Texas Center for Superconductivity and Department of Physics, University of Houston, Houston, Texas 77204, USA\\
$^{9}$Department of Information Systems and Cybersecurity, Grove School of Engineering, The City College of New York, 275 Convent Avenue, New York, New York 10031, USA
}

\date{\today}

\begin{abstract}

We have obtained and analyzed the electronic states, polarization function and the plasmon excitations for $\alpha-\mc{T}_3$-based
nanoribbons with armchair termination. The calculated plasmon dispersions strongly depend on the number of the atomic rows across the
ribbon, and the presence of the energy gap between the valence and conduction bands which is also determined by the nanoribbon
geometry. The bandgap was proven to have the strongest effect on both the plasmon dispersions and their Landau damping. We 
have also demonstrated that for a small electron doping the plasmon dispersions do not depend on the relative hopping parameter
$\alpha$ of the considered $\alpha-\mc{T}_3$ material in the long-wave limit and investigated the conditions when $\alpha$ 
becomes an important factor which strongly affects the plasmons. We believe that our new uncovered electronic and collective
properties of nano-size $\alpha-\mc{T}_3$ ribbons will find their applications in the field of modern electronics and nanodevices.
\end{abstract}

\maketitle

\section{Introduction}
\label{s01}

Graphene plasmons, or the quantum collective oscillation of its free electrons has become one of the most important and quickly growing 
fields in connection with the unique electronic and optical properties of all recently discovered Dirac cone materials, graphene\,\cite{grigorenko2012graphene} and beyond. The plasmons are viewed a crucial instrument in optical manipulation, light sensing, nanoscale spectroscopy and other applications.\,\cite{garcia2014graphene} The plasmon-based optical devices demonstrated good efficiency in various frequencies, specifically in terahertz with a possibility to move them up to the visible light range by decreasing the size of a graphene sample to a nanoribbon (GNR). 

\par 
One of the most crucial problems in connection with the graphene plasmonic applications is a precise estimate of plasmon frequencies or the energy range, regions of finite (or negligible) Landau damping where a plasmon is reduced into single particle excitations and its behavior in the long wave limit. After more than a decade of extensive studies,\,\cite{hwang2007dielectric,pyatkovskiy2008dynamical,wunsch2006dynamical,
iurov2017thermal, gumbs2014strongly, malpd, iurov2017controlling, gumbs2015tunable} modern many-body theory of low-dimensional materials has developed a huge arsenal of the tools of plasmon investigation, primarily based on calculating the dynamical polarization function,\,\cite{gonccalves2016introduction} which is also related to the static screening transport properties of the investigated material.

\par
\medskip 
\par

Graphene nanoribbons (GNR's), or nano-size strips made from two-dimensional graphene, has become one of the hotspots in low-dimensional electronic because of their specific size and shape which are convenient and suggestive of their use in nanoelectronic devices. They also reveal a number of spectacular and technologically promising properties, as well as some unique physical phenomena which had not been 
observed in the corresponding bulk materials.\,\cite{groning2018engineering} These includes exotic and non-trivial topological electronic
states, (Majorana fermions)\,\cite{cao2017topological, kitaev2001unpaired}, spin–momentum correlation and locked transport channels,  
arrays of plasmonic nanoantennas \cite{pfeiffer2018enhanced} Specific electronic quantum phases could be created at the 
junctions of armchair nanoribbons.\,\cite{sevinccli2008superlattice} Transport of charged carriers was studied in the networks of armchair nanoribbons and possibility of a reproducible field-effect transistor with higher field-effect mobilities was also demonstrated \,\cite{richter2020charge}

\par 
\medskip 
\par 
Plasmonics has become on of the central areas in low-dimensional device physics since the collective excitations 
could be confined inside the patterned ribbons which results in a distinct plasmon mode and strong enhancement of the external
field.\,\cite{fei2015edge, karimi2017plasmons, xia2016localized} Specific types of plasmon excitations in NR's \,\cite{gomez2016plasmon} have a number of important and sometimes unexpected applications in sensing and nano-imaging.\,\cite{khaliji2020plasmonic, hu2017imaging, xia2018plasmonically}

\par 
All the electronic properties of nanoribbons strongly depend on their size (atomically precise width of a ribbon) and 
specifically, on the type of the termination \,\cite{f2, andersen2012plasmon} - zigzag\,\cite{hancock2010generalized, rodrigues2011zigzag} or 
armchair.\,\cite{raza2008armchair, kimouche2015ultra, zheng2007analytical} Thus, the investigation of quantum finite-size effects and nonlocality on the plasmons, dielectric and optical response in nanoribbons \,\cite{zhao2020optical} and nanodisks demonstrated substantial plasmon broadening which is significantly larger for zigzag termination compared to the armchair case.\,\cite{thongrattanasiri2012quantum} A tunable band gap in the which is not generally present in graphene and could be only introduced in a limited range by applying an external field \,\cite{kiMain} is essential for most of the semiconductor devices, directly depends on the width of an armchair nanoribbon \,\cite{zheng2007analytical, rozhkov2009electronic, andersen2012plasmon} and, therefore, could be set by creating and using an AGNR with a specific number of atomic cells across the ribbon.

\par 

It is important that efficient reliable and affordable techniques of the fabrication of nano-size ribbons with a given 
width have been developed using chemical vapor deposition, \,\cite{chen2016synthesis, zhang2004structure,talirz2016surface}
in addition to earlier existing atomically precise bottom-up fabrication \,\cite{xu2016recent} based on the chemical or lithographic 
unzipping of carbon nanotubes,\,\cite{cai2010atomically} which made it possible to consider the electronic properties of a nanoribbon with a fixed and precise atomic width.

\par 
\medskip
\par 

Among all the newest and recently discovered two dimensional structures,\,\cite{liu2016van} $\alpha-\mc{T}_3$ model represents one of the most unusual and budding materials. Its atomic building represents a hexagonal honeycomb lattice like we observed in graphene with an additional atom located  at the center of each hexagon - a hub (or $H$)atom. The interaction strength and the resulting electron hopping integral between $H$ and the remaining $A$ and $B$ rim atoms differ from such hopping coefficient  between the nearest-neighbor rim atoms of hexagon.
The relative hopping parameter $\alpha = t_{hub-rim}/t_{rim-rim}$ could vary between 0 and 1. Its lowest value $\alpha = 0$  corresponds to
graphene with a completely detached set of hub atoms, and the opposite limit $\alpha = 1$ is defined as a dice lattice. A general $\alpha-\mc{T}_3$ model is considered an interpolation between graphene and a dice lattice. A number of really existing at3 materials has been successfully fabricated.\,\cite{Add1} 

\medskip
Such an atomic structure results in a pseudospin-1 Dirac-Weyl Hamiltonian and the metallic (gapless) low-energy bandstructure which consists of a Dirac cone and an additional dispersionless flat band. The flat band makes the $\alpha-\mc{T}_3$ distinguished from any other Dirac materials and appears to be very stable and robust in the presence of external fields\,\cite{ourpeculiar, dey1, dey2} or a disorder. A energy 
bandgap could be also generated in a dice lattice \,\cite{gorb} similarly to how it was done in graphene. Recently, Dirac semifinals also demonstrates some interesting electronic phenomena \,\cite{isl1, isl2} close but not completely similar to $\alpha-\mc{T}_3$. The unique electron bandstructure of $\alpha-\mc{T}_3$ model leads to its unusual electronic, optical, collective, magnetic and  
topological properties \,\cite{
berc,
weekes2021generalized,
wu2021superfluid,
iurov2020quantum,
opt1,
yeke,
anwar2020interplay,
alphaDice, alphaKlein, iurov2020klein,
wa20,
Dutta1,
zhou2021andreev,
mojarro2020electron,
iurov2020many,
tutul1, tutul2,
nic1, nic2, thesis} which has been rigorously investigated over the last several years.

\par 
\medskip 
\par 
In spite of the fact that $\alpha-\mc{T}_3$ materials were discovered very recently, there have been a handful of crucial publications 
on the subject of the nanoribbons made from such materials. The group velocities and current distributions were also studied in such ribbons with  both armchair and zigzag termination were investigated in Ref.~[\onlinecite{last}] in the presence of magnetic field. A comprehensive study in the electronic states in a dice lattice with $\alpha = 1$ demonstrated new electronic states without a direct analogy in graphene nanoribbon for the case of zigzag edges.\,\cite{gus_main}  The corresponding electronic states in the presence of magnetic field was performed in Ref.~[\onlinecite{bugaiko2019electronic}] for both armchair and zigzag terminations. A mean-field investigation of the strain effect revealed a transition antiferromagnetic to ferromagnetic with increasing $\alpha$ \,\cite{Cheng_2021} in analogy with paramagnetic transition 
in a bulk $\alpha-\mc{T}_3$.\,\cite{piech1, piech2} A valley degree of freedom and lifting the valley degeneracy is also of the highest interest for both bulk\,\cite{ourpeculiar} and nanoribbons. \,\cite{tan2020valley} Ferromagnetic ordering in dice ribbons was investigated in Ref.~[\onlinecite{soni2020flat}]

\par 
\medskip 
\par 
Most of the existing theoretical papers on GNR plasmonics deal with a very specific case of semi-metallic gapless dispersions and low electron doping for which an approximated analytical derivation of the long-wave plasmon dispersions could be done, \,\cite{f2, mr1, mr2} 
which are realistic and important but very limited because the Dirac Hamiltonian and $k \cdot p$ approximation works well only for the nanoribbons with sufficiently large width in which even for a moderated doping density several subbands become populated. A crucial case of a finite bandgap has never received sufficient attention and consideration even for earlier investigated graphene nanoribbons. In view of this, our primary focus for this paper is to investigate the plasmons, their dispersions and damping for the various non-trivial cases of nanoribbon widths which determine the gap in the single-particle dispersions, finite-level electron doping densities and, specifically, the dependence of the obtained plasmon excitations on the relative hopping parameter $\alpha$. 

\par 
\medskip 
\par 
The rest of this paper is organized in the following way. In Sec.~\ref{s02}, we derive the phase-dependent electronic states and demonstrate that the energy dispersions are the same for graphene, dice lattice and all kinds of $\alpha$-$T_3$ materials, i.e, do not depend on $\alpha$. The polarization function, dielectric function matrix elements, Coulomb potential and the plasmon dispersions are obtained in Sec.~\ref{s03} where we also provide a brief description of the different wave function overlaps, their dependence on the the transfer wave vector $q$ and parameter $\alpha$. Our numerically obtained plasmon dispersion, as well as some crucial analytical expression for the long-wave limit are presented and analyzed in Sec.~\ref{s04}. Finally, Sec.~\ref{s05} contains some concluding remarks, possible applications of our results and further research outlook.

\section{$\alpha$-dependent electronic states in finite-width nanoribbons}
\label{s02}

\begin{figure}
 \centering
 \includegraphics[width=0.9\linewidth]{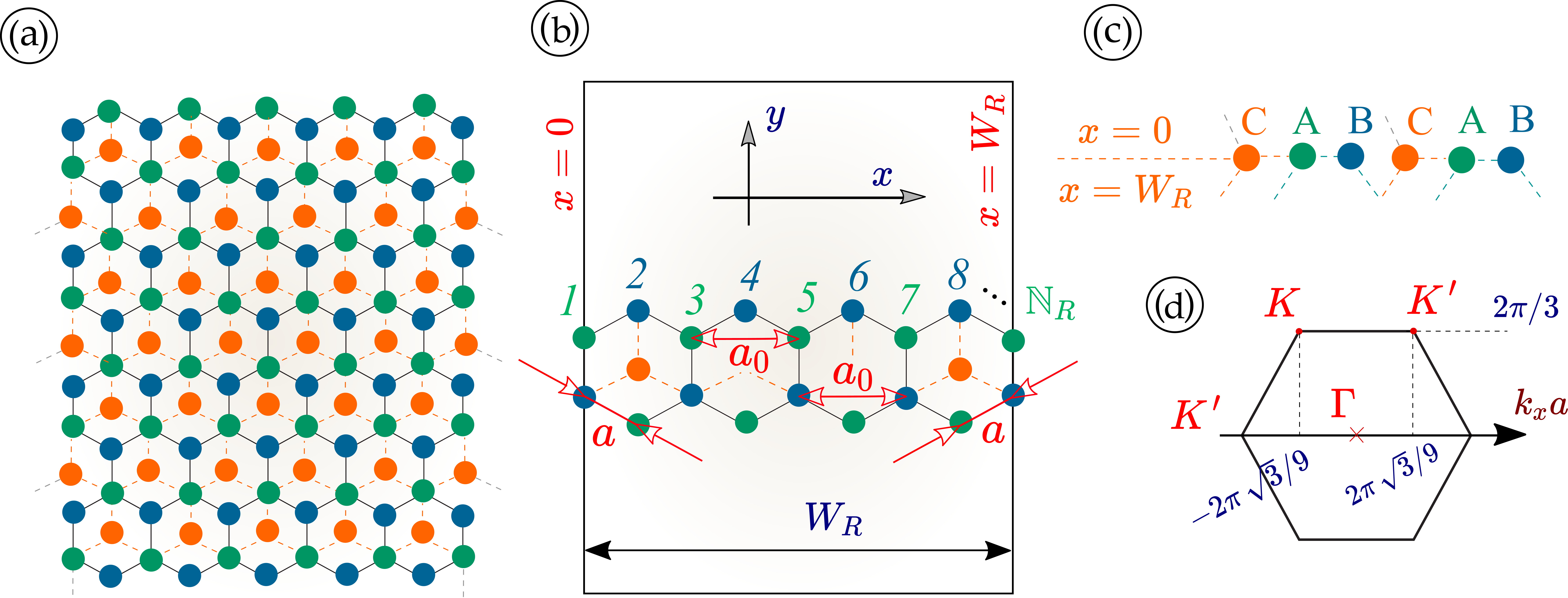}
\caption{Armchair nanoribbons made from a two-dimensional $\alpha - \mc{T}_3$ material. $(a)$ and $(b)$ shows the schematics for a nanoribbon with armchair termination, calculating the width of such a ribbon and define atomic distance $a$ and lattice constant $a_0$. Panel $(c)$ demonstrates the edge row and the boundary conditions for an armchair margin in $\alpha - \mc{T}_3$ or a dice lattice. Plot $(b)$  
shows the definition of the wave vectors for $K$ and $K'$ valleys. 
}
\label{FIG:1}
\end{figure}

Our goal now is to calculate the wave function and the corresponding low-energy electron dispersions in a nanoribbon 
made from an $\alpha - \mc{T}_3$ material described by a bulk pseudospin-$1$ Dirac-Weyl Hamiltonian

\begin{equation}
\label{Hbulk}
 \mathbb{H}_b^{\tau,\,\phi}({\bf k}) = \gamma_0 \left\{
  \begin{array}{ccc}
   0 & k^\tau_- \,   \cos \phi & 0 \\
    k^\tau_+ \, \cos \phi & 0 & \,  k^\tau_- \, \sin \phi   \\
   0 & k^\tau_+ \, \sin \phi  & 0
  \end{array}
 \right\}) \, ,
\end{equation}
where $k^\tau_\pm = \tau k_x \pm i k_y$ depends on the valley index $\tau$, i.e., is not the same for the two non-equivalent $K$ and $K'$ valleys. We limit our consideration to the low-energy states located in the vicinity of the valleys with $\delta K_x = (4 \pi)(a_0)\,1/(3 
\sqrt{3})$, $\delta K_y = 0$. Relative hopping parameter $\alpha$ is related to the phase $\phi$ which is present in Eq.~\eqref{Hbulk} (which is sometimes is also referred as Berry phase, even though the Berry phase in $\alpha - \mc{T}_3$ has been obtained as 
$\pm \cos(2\phi)$) \,\cite{illes2015hall, ourpeculiar} as $\alpha = \tan \phi$. Distance $a_0 = 0.142\,mn$ is the lattice constant (distance between the nearest identical atoms, such as $B$ and $B$) and $a = a_0/(2 \sin 30^0) = a_0/\sqrt{3}$ is the side of a hexagon in a lattice, as 
shown in Fig.~\ref{FIG:1}. 

\medskip 
In a bulk $\alpha - \mc{T}_3$ model, the three solutions for the low-energy band structure $\varepsilon^{\gamma = \pm 1}_{\tau, \, \phi}({\bf k}) = \pm \gamma \hbar v_F k$ with $\gamma = \pm 1$ correspond to the valence and conduction bands and are exactly similar to graphene. Apart from these two bands, Hamiltonian \eqref{Hbulk} allows for an additional solution to $\varepsilon^{\gamma = \pm 1}_{\tau, \, \phi}({\bf k}) = 0$ which represents a dispersionless or flat band. This general schematics with the division on the valence, conduction and the flat bands is also observed in nanoribbons. All the energy dispersions do not directly depend on the valley index $\tau$ in contrast to the phases of the corresponding electronic states.

\medskip 
The finite width and the edge termination of a ribbon establishes its all crucial electronic properties such as the quantization of the transverse electron momentum and possible existence of a bandgap in the electronic dispersions. The width of a ribbon $W_R$ is related to 
the number of the atomic rows $\mbb{N}_R$ across the ribbon as $W_R = a_0/2 (\mbb{N}_R-1) = a \sqrt{3} \, (\mbb{N}_R-1)/2$, such as shown in Fig.~\ref{FIG:1}. Here, $\mbb{N}_R$ is the total number of all atomic rows including all types ($A$, $B$ and $H$) of the lattice
atoms.

\medskip

The case of a finite-width nanoribbon is distinguished primarily because there is no translational symmetry in the transverse ($x-$) 
directions so that $k_y \longrightarrow - i \pr/\pr x$ and Hamiltonian \eqref{Hbulk} is modified as     

\begin{equation}
\label{Hrib}
 \mathbb{H}_r^{\tau,\,\phi}({\bf k}) = \gamma_0 \left\{
  \begin{array}{ccc}
   0 & (-i \tau \pr/\pr x - i k_y) \,   \cos \phi & 0 \\
    (-i \tau \pr/\pr x + i k_y)  \, \cos \phi & 0 & \,  (-i \tau \pr/\pr x - i k_y)  \, \sin \phi   \\
   0 & (-i \tau \pr/\pr x + i k_y)  \, \sin \phi  & 0
  \end{array}
 \right\} \, ,
\end{equation}
which basically means that transverse momentum $k_x$ cannot be introduced as a good quantum number. 
 
\medskip
The boundary conditions are obviously the same as in dice lattice \,\cite{gus_main} and reflect the fact that 
all three sublattice wave function, including the $H$ atom, disappear at each boundary of our ribbon

\begin{eqnarray}
\label{bcon}
&& \varphi_{\nu}(x)\big|_{x=0} = \varphi_{\nu}(x)\big|_{x=0}  \, , \\
\nonumber
&& \varphi_{\nu}(x)\big|_{x=L} = \tet{exp}(i \delta K_x \, W_R) \, \varphi_{\nu}(x)\big|_{x=L} \, ,
\end{eqnarray}
where $\nu = A$, $B$ and $H$. We see that boundary conditions \eqref{bcon} contain the states for both $K$ and $K$' valleys therefore, 
these electronic states are mixed similarly to graphene, \,\cite{f2} and both of the valleys needs to be taken into our consideration. 

\medskip 
Therefore, the complete wavefunctions are

\begin{equation}
\label{wave0}
\Phi^\phi_{\gamma}(n \, \vert \, x, k_y) = \left\{ 
\begin{array}{c}
\Psi^{\tau= 1 ({\bf K}), \phi}_{\gamma}(n \, \vert \, x, k_y) \\
\Psi^{\tau= - 1 ({\bf K})', \phi}_{\gamma}(n \, \vert \, x, k_y) 
\end{array}
\right\} \, .
\end{equation}
Each of the two components in \eqref{wave0} is related to a separate valley ($\tau = \pm 1$) and is built as 

\begin{equation}
\label{wave01}
\Psi^{{\bf K}, \phi}_{\gamma}(n \, \vert \, x, k_y) = \left\{ 
\begin{array}{c}
\varphi_A(x \, \vert \, \phi, n) \\
\varphi_H(x \, \vert \, \phi, n) \\ 
\varphi_B(x \, \vert \, \phi, n)
\end{array}
\right\} \, \tet{e}^{i k_y y}
\end{equation}
and 
\begin{equation}
\label{wave02}
\Psi^{{\bf K}', \phi}_{\gamma}(n \, \vert \, x, k_y) = \left\{ 
\begin{array}{c}
\varphi_B'(x \, \vert \, \phi, n) \\
\varphi_H'(x \, \vert \, \phi, n) \\
\varphi_A'(x \, \vert \, \phi, n)
\end{array}
\right\} \, \tet{e}^{i k_y y} \, . 
\end{equation}

The unknown components of eigenstates \eqref{wave01} and \eqref{wave02} could be now obtained from Hamiltonian \eqref{Hrib}. Specifically, the hub state $\varphi_H (x)$ is given by \,\cite{gus_main}

\begin{equation}
\label{phiH}
\frac{1}{2} \, k_\varepsilon^2 \, \varphi_H (x) = - \left[ 
\frac{\pr^2}{\pr\,x^2} - k_y
\right] \varphi_H (x) \, , 
\end{equation}
with the following general solution  

\begin{equation}
\label{phiHs}
\varphi_H (x) = A \, \tet{e}^{i \xi x} + B \, \tet{e}^{- i \xi x} \, . 
\end{equation}

While both terms in solution \eqref{phiHs} are relevant and could be a part of the sought wave function, we also keep in mind that the 
wave function corresponding to each valley has only one direction of the transverse momentum $\xi$  ($\backsim \tet{e}^{+ i \xi x}$ for $K$ and $\backsim \tet{e}^{- i \xi x}$ for $K'$) and assume $\varphi_H (x) = A \, \tet{e}^{i \xi x}$ and $\varphi_H' (x) = A' \, \tet{e}^{-i \xi x}$. The boundary conditions for $\varphi_H^{(')} (x)$ are as follows: 

\begin{eqnarray}
\label{sys0}
&& A - B' = 0 \\
\nonumber 
&& A \, \tet{e}^{i \xi W_R} - \tet{e}^{i \delta K_x \, W_R} \, B' \, \tet{e}^{- i \xi W_R} = 0 \, . 
\end{eqnarray}
System \eqref{sys0} has non-trivial solutions if 

\begin{equation}
\label{det01}
\tet{e}^{i \xi W_R} - \tet{exp}[i \delta K_x \, W_R] \, \tet{e}^{- i \xi W_R} = 0 \, .
\end{equation}
Since $W_R$ could be chosen arbitrarily, equation \eqref{det01} is equivalent to

\begin{eqnarray}
\label{xiN1}
&& \xi W_R = 2 \pi N + \delta K_x \, W_R - \xi W_R \, ,  \\
\nonumber
&& \text{where} \hskip0.2in \, N = 0, \pm 1, \pm 2, \pm 3, ... \, . 
\end{eqnarray}

Condition \eqref{xiN1} which determines the quantization of the transverse electron momentum $\xi_N$ and the complete energies of our states, is equivalent to what was earlier found for both limiting cases of graphene ($\alpha = 0$) \,\cite{f1,f2} as a dice lattice 
($\alpha = 1$).\,\cite{gus_main} 

\par 
The obtained result in \eqref{xiN} also reflects the fact that the energy dispersions $\varepsilon_n(k_y)$ are the same in graphene and 
$\alpha - \mc{T}_3$ materials, except for the presence of a flat band. This is similar to what we earlier observed for the corresponding 
bulk materials.

\par 
From Eq.~\eqref{xiN} we conclude that the transverse electron momentum is quantized as 

\begin{equation}
\xi_N = \frac{\pi N}{W_R} - \frac{4 \pi}{3 a_0} = \frac{2 \pi}{\sqrt{3}\,a} \,\left(  
\frac{N}{\mbb{N}_R+1}- \frac{1}{3}
\right)  = \frac{2 \pi}{3 \sqrt{3}\,a} \,
\frac{3 N - \mbb{N}_R - 1}{\mbb{N}_R+1}
\, ,
\end{equation}

The energy dispersions $\varepsilon_{\,n}(k_y) = \gamma_0 \, \xi^{\,\sigma=0}(k_y)$ are now given by 

\begin{equation}
\varepsilon^{\sigma = \pm 1}_{\,N}(k_y) = \pm \sqrt{
k_y^2 + \xi_{\,N}(W_R)^{\,2}  
} = 
\pm \left\{ 
k_y^2 + \left(
 \frac{\pi N}{W_R} - \frac{4 \pi}{3\sqrt{3} \,a_0}
\right)^2 \,
\right\}^{1/2}
\, .
\end{equation}

\begin{figure}
 \centering
 \includegraphics[width=0.55\linewidth]{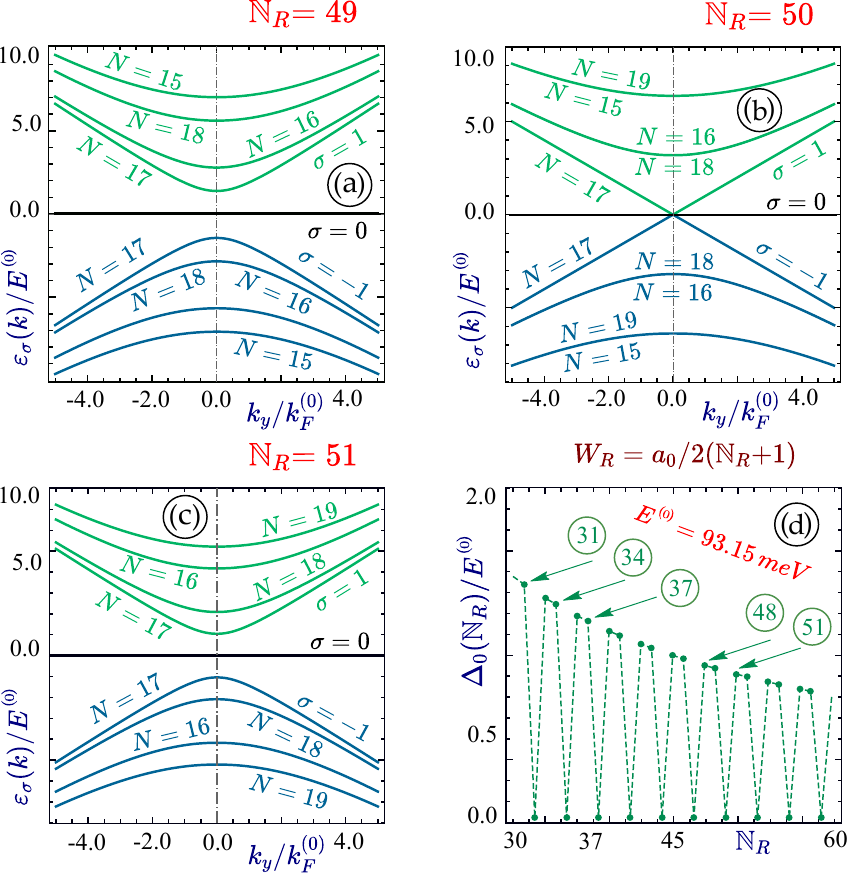}
\caption{(Color online) Low-energy subbands for an $\alpha-\mc{T}_3$ armchair nanoribbons of different widths. Panels $(a)$, $(b)$ and 
$(c)$ demonstrate the energy dispersions as a function of the longitudinal momentum $k_y$ for the ribbons with $\mbb{N}_R = 49$, $50$ and 
$51$ atomic rows, correspondingly. Plot $(d)$ shows the energy bandgap $\Delta_0(\mathbb{N}_R)$ between the conduction and flat bands in the nanoribbons with various number of atomic rows $\mathbb{N}_R$. 
}
\label{FIG:2}
\end{figure}

\medskip
The obtained energy dispersions presented in Fig.~\ref{FIG:2}, reflect a known fact that depending on the number $\mbb{N}_R$
there might be either zero or finite bandgap. The gap between the valence (or conduction) and the flat band which is a half of the total bandgap is

\begin{equation}
\Delta_0(\mbb{N}_R) = \frac{4 \pi}{3 \sqrt{3}\,a} \, \frac{\gamma_0}{\mbb{N}_R +1} \, \text{Min}_N (3 N - \mbb{N}_R  - 1)
 \, ,  
\end{equation}

where the minimal value of $3 N - \mbb{N}_R  - 1$ for a given ribbon width $\mbb{N}_R$ is achieved for the integer $N^{\Delta}$
which is obviously zero if $\mbb{N}_R  + 1$ is exactly divisible by 3. The obtained number $N_0 = (\mbb{N}_R  + 1)/3$ specifies the 
lowest pair of metallic subbands which touch each other and the flat band at Dirac point. If this is not the case, the number of the
two lowest subbands $N^{\Delta}$ obtained as the least of the following two numbers: 

\begin{equation}
N^{\Delta}_{(1,2)}= \text{InP} \left( \frac{\mbb{N}_R  + 1}{3}\right) \pm 1 \, , 
\end{equation}
where $\text{InP}(x)$ means the integer part of a rational number $x$. The dependence of the energy gap vs. the ribbon width $\mbb{N}_R$ is presented in Fig.~\ref{FIG:2} $(d)$. All the other (higher) subbands of the metallic dispersions are doubly degenerate, as we show in the remaining panels of Fig.~\ref{FIG:2}. We also see that energy separation and the gap are large for a narrow ribbon, similarly to the case of 
a quantum well, which makes it difficult to populate more than one subband for an experimentally accessible electron doping density. 

\medskip 
In all our calculations, the unit of momentum is chosen $k_F^{(0)} = \pi/2 (10^8 \, m^{-1}) = 1.57 \cdot 10^8 \,m^{-1}$, while the unit of length is its reciprocal $L_0 = 1/k_F^{(0)} = 6.33\,nm$. As for the energy, its unit is $E^{(0)} = \gamma_0 k_F^{(0)} = 93.15\,meV$ which corresponds to a plasmon frequency $1.42 \cdot 10^{14}\,Hz$. The width of a ribbon is estimated as $W_R = 6.27\,nm$ for $\mathbb{N}_R = 50$ and $24.72\,nm$ for $\mathbb{N}_R = 200$.  

\medskip
Once $\varphi_H (x)$ is known, we can immediately find the remaining components of the wave functions from 

\begin{eqnarray}
&& \varphi_A^{(')}(x) = \frac{1}{k_\varepsilon} \, \left(  
- i \tau \frac{\pr}{\pr x} - i k_y 
\right) \, \varphi_H^{(')}(x)
\, ,  \\
\nonumber
&& \varphi_C^{(')}(x) = \frac{1}{k_\varepsilon} \, \left(  
- i \tau \frac{\pr}{\pr x} + i k_y 
\right) \, \varphi_H^{(')}(x)
\, . 
\end{eqnarray}

The obtained expressions are as follows: 

\begin{equation}
\Psi^{\tau, \phi}_{\sigma = \pm 1}(n \, \vert \, x, k_y) = \frac{1}{\sqrt{W_R}} \, 
\varphi^{\tau, \phi}_{\sigma = \pm 1}(n \, \vert \, x) \, \tet{e}^{i k_y y} \, , 
\end{equation}

where

\begin{equation}
\label{wave11}
\varphi^{\tau, \phi}_{\sigma = \pm 1}(n \, \vert \, x) = \frac{1}{\sqrt{2}}
\left\{
\begin{array}{c}
\tau \cos \phi \, \tet{e}^{- i \tau \Theta_n}  \\
\sigma \\
\tau \sin \phi \, \tet{e}^{i \tau \Theta_n}
\end{array}
\right\} \, . 
\, \tet{exp}[i \tau \xi_n x]
\end{equation}

It is crucial to realize that only the electron/hole index $\sigma = \pm 1$ and not the subband number $N$ determines the  $\pm$ sign of the energy dispersions $\epsilon^\sigma_N (k_y)$. 

\medskip 
Similarly to a bulk $\alpha-mc{T}_3$, a nanoribbon is expected to exhibit a zero-energy flat band because equation \eqref{phiH} allows for an additional solutions with $\varepsilon(k) = 0$

\begin{equation}
\left[
 \frac{\pr}{\pr x} - \tau k_y 
\right] \, \cos \phi \, \varphi_A^{(')}(x) +
\left[
\frac{\pr}{\pr x} + \tau k_y 
\right] \, \sin \phi \, \varphi_B^{(')}(x) = 0
\end{equation}

or, explicitly, 

\begin{eqnarray}
\label{s1}
&& \left[
\frac{\pr}{\pr x} - k_y 
\right] \, \cos \phi \, \varphi_A(x) +
\left[
\frac{\pr}{\pr x} + k_y 
\right] \, \sin \phi \, \varphi_B(x) = 0 \, , \\
\nonumber 
&& \left[
 \frac{\pr}{\pr x} + k_y 
\right] \, \cos \phi \, \varphi_A'(x) +
\left[
\frac{\pr}{\pr x} - k_y 
\right] \, \sin \phi \, \varphi_B'(x) = 0 \, , 
\end{eqnarray}

from which we see that the two equations for $K$ and $K$' valley differ only by the sign of x-derivative $\pr/\pr x$. 

\par 
\medskip 

The two equations \eqref{s1} have an obvious solutions in the form of $\varphi_\nu \backsim \tet{e}^{\pm i \xi x}$. However, we 
also keep in mind that the wave function corresponding to each valley has only one direction of the transverse momentum $\xi$ 
($\backsim \tet{e}^{+ i \xi x}$ for $K$ and $\backsim \tet{e}^{- i \xi x}$ for $K'$).

\par 
Therefore, we assume $\varphi_A(x) = A \tet{e}^{i \zeta \,x}$, $\varphi_B(x) = B \tet{e}^{i \zeta \,x}$, and  
$\varphi_A'(x) = A' \tet{e}^{i \zeta \,x}$ and $\varphi_B'(x) = B' \tet{e}^{i \zeta \,x}$, correspondingly. 
Eqs.~\eqref{s1} could be now rewritten as 

\begin{eqnarray}
\label{s2}
&& \cos \phi \, \left( 
- i \zeta + k_y
\right) \, A +
\sin \phi \, \left( 
- i \zeta + k_y
\right) \, B = 0 \, , \\
\nonumber
&& \cos \phi \, \left( 
+ i \zeta - k_y
\right) \, A' +
\sin \phi \, \left( 
+ i \zeta + k_y
\right) \, B' = 0 \, ,
\end{eqnarray}

where we took into account that Eqs.~\eqref{s1} should be satisfied for all $x$: $0 < x < W_R$. Another four equations come from
the boundary conditions \eqref{bcon}:

\begin{eqnarray}
\label{s31}
&& A - A' = 0 \, , \\
\label{s32}
&& B - B' = 0 \, , \\
\label{s33}
&& A \, \tet{e}^{i \xi W_R} - \tet{e}^{i \delta K_x \, W_R} \, A' \, \tet{e}^{- i \xi W_R} = 0 \, , \\
\label{s34}
&& B \, \tet{e}^{i \xi W_R} - \tet{e}^{i \delta K_x \, W_R} \, B' \, \tet{e}^{- i \xi W_R} = 0 \, ,
\end{eqnarray}

which obviously makes our system overdetermined. Eqs.~\eqref{s2}, \eqref{s31} and \eqref{s33} are compatible and have 
a non-zero solution if  

\begin{equation}
\label{det1}
- \sin^2 \phi \,\left\{ 
\tet{e}^{i \xi W_R} - \tet{exp}[i \delta K_x \, W_R] \, \tet{e}^{- i \xi W_R}
\right\} \, \left(  
k_y^2 + \xi^2
\right) = 0
\end{equation}

is satisfied. Since $k_y$ could be chosen arbitrary, this is equivalent to the following condition 

\begin{eqnarray}
&& \xi W_R = 2 \pi N + \delta K_x \, W_R - i \xi W_R \, ,  \\
\nonumber
&& \text{where} \hskip0.2in \, N = 0, \pm 1, \pm 2, \pm 3, ... \, . 
\end{eqnarray}

or, equivalently,

\begin{equation}
\label{xiN}
\xi_N = \frac{\pi N}{W_R} + \frac{1}{2} \, \delta K_x \, .
\end{equation}

\par
\medskip

Now, the wavefunctions for the flat band with $\gamma=0$ could be immediately obtained as

\begin{equation}
\Psi^{\tau, \phi}_{\gamma = 0}(n \, \vert \, x, k_y) = \frac{1}{\sqrt{W_R}} \, 
\varphi^{\tau, \phi}_{\gamma = 0}(n \, \vert \, x) \, \tet{e}^{i k_y y}
\end{equation}

where 

\begin{equation}
\label{Wave01}
 \varphi^{\tau, \,\phi}_{\gamma = 0}(n \, \vert \, x) =  \left\{
 \begin{array}{c}
 \sin \phi \,\, \left(\tau \, \xi_{n} - i k_y \right)  \\
 0 \\
 -\cos \phi \,\, \left(\tau \,  \xi_{n} + i k_y \right)
 \end{array}
 \right\} \,
\frac{\tet{e}^{i \tau \xi_{n} \, x}}{k_\varepsilon} \, ,
\end{equation}

in which the $K$ and $K'$ valleys correspond to $\tau = \pm 1$. If we introduce the following notation $\Theta_n =\Theta_{(\xi_n,k_y)}$ 
is the angle associated with the quantized wave vector ${\bf k}_n=\{\xi_n,k_y\}$ so that $\Theta_n = \tan^{-1}(k_y/\xi_n)$, the 
result in Eq.~\eqref{Wave01} could be simplified as 

\begin{equation}
\label{Wave02}
\varphi^{\tau, \phi}_{\gamma = 0}(n \, \vert \, x) = 
\left\{
\begin{array}{c}
\tau \sin \phi \, \tet{e}^{- i \tau \Theta_n}  \\
0 \\
- \tau \cos \phi \, \tet{e}^{i \tau \Theta_n}
\end{array}
\right\}
\, \tet{exp}[i \tau \xi_n \, x] \, . 
\end{equation}

Finalizing the discussion of the obtained electronic states, we should say that the type, structure and dependence of the wave functions
\eqref{wave11} and \eqref{Wave02} are similar to those in a bulk material. However, the transverse momenta $\xi_n$ and the quantum phase 
$\Theta_n$ are now quantized (discrete), and the corresponding quantization rules \eqref{xiN} account for the mixing of $K$ and $K$' valleys.

\begin{figure}
 \centering
 \includegraphics[width=0.55\linewidth]{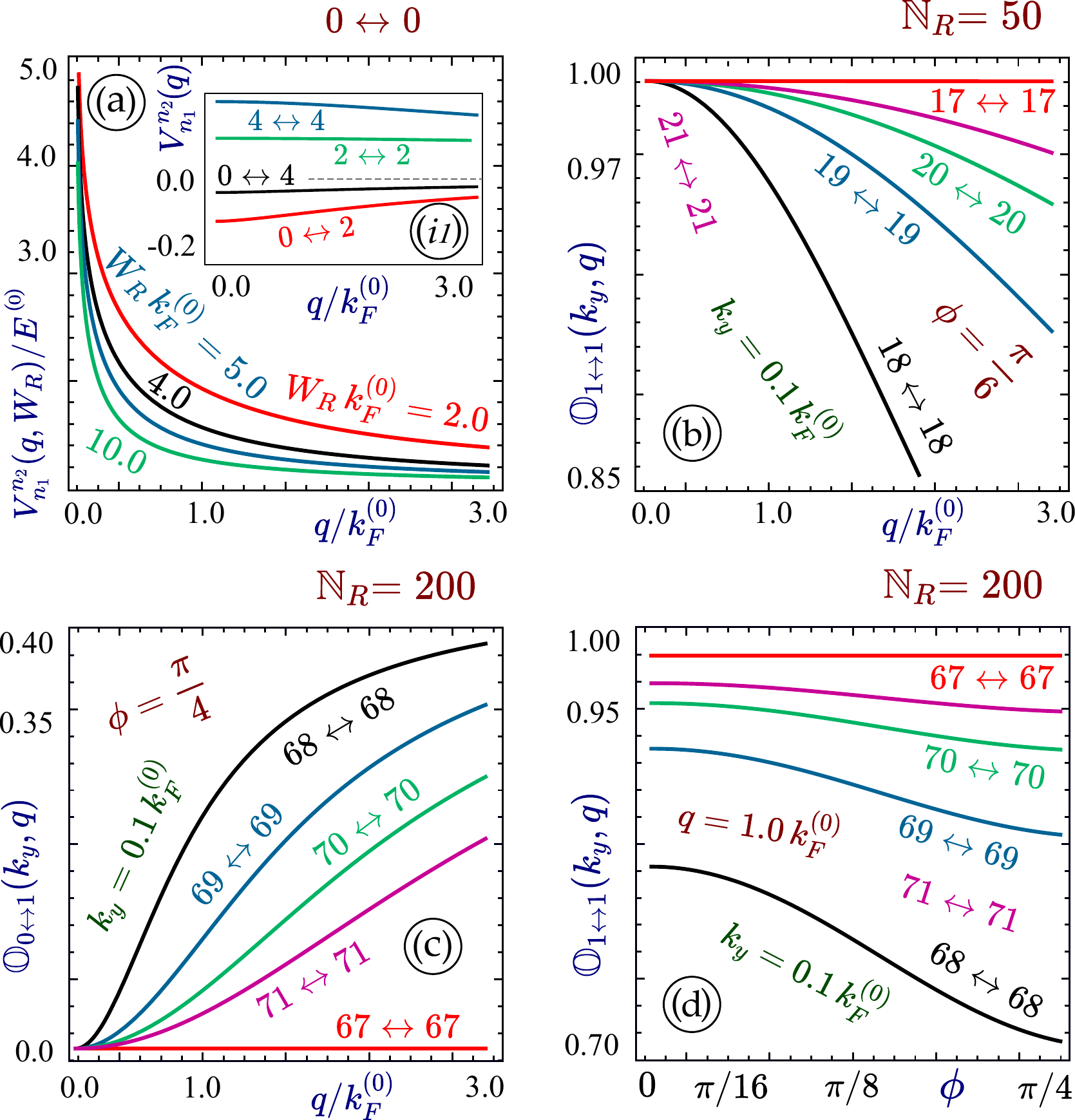}
\caption{(Color online) Coulomb potential and wave function overlaps for nanoribbons with armchair termination. $(a)$ represents various matrix elements $V_{n_1}^{n_2}(q W_R)$ of the Coulomb potential with $n_1 = i-j = 0$ and $n_2 = m-n = 0$ as a function of the transfer wave vector $q$ for the different widths, and inset $(i1)$ shows the remaining elements with $n_1 \neq 0$ or $n_2 \neq 0$, as labeled. $(b)$, $(c)$ and $(d)$ demonstrate \textit{intra-subband} ($n_1 = n_2$) overlaps $\mathbb{O}_{\sigma_1 \leftrightarrow \sigma_2}^{n_1,n_1}(k_y,q)$ for the various types of \textit{inter-} $0 \leftrightarrow 1$ and \textit{intra-band} $1 \leftrightarrow 1$ electron transition as a function of $q$. The number of atomic rows $\mbb{N}_R$ across the ribbon was chosen as $50$, $50$ and $200$ for panels $(b)$, $(c)$ and $(d)$, correspondingly.       
}
\label{FIG:3}
\end{figure}

\section{Polarization function, plasmon dispersions and damping}
\label{s03}

\begin{figure}
 \centering
 \includegraphics[width=0.6\linewidth]{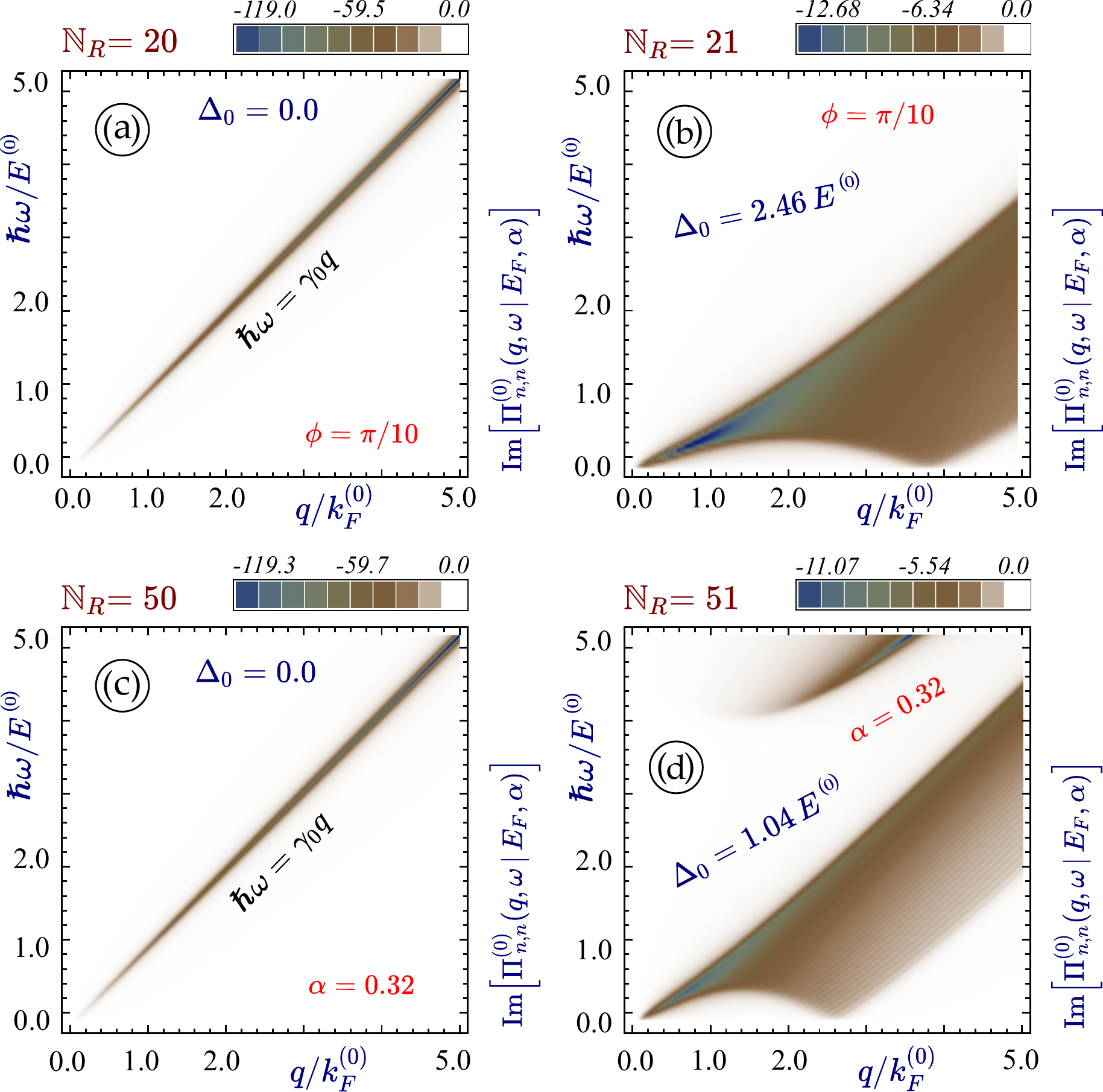}
\caption{(Color online) Single-particle excitation spectrum for nanoribbons with different widths. All panels $(a)$-$(d)$ represent 
the density plots of the imaginary part of the non-interacting \textit{intra-subband} polarization functions $\Pi^{(0)}_{n,n'}(q, \omega \, \vert \, E_F, \alpha)$ which depend on frequency $\omega$ and wave vector $q$ for $\mbb{N}_R = 20$, $21$, $50$ and $51$, correspondingly. The energy bandstructure for $\mbb{N}_R = 20$ and $\mbb{N}_R = 50$ (plots $(a)$ and $(c)$) is metallic, while the other two cases shown in plots $(b)$ and $(d)$ demonstrate finite energy bandgaps between the valence and conduction bands. Relative hopping parameter $\alpha$ was selected $0.32$ (phase $\phi=\pi/10$) for all plots.         
}
\label{FIG:4}
\end{figure}

\begin{figure}
 \centering
 \includegraphics[width=0.6\linewidth]{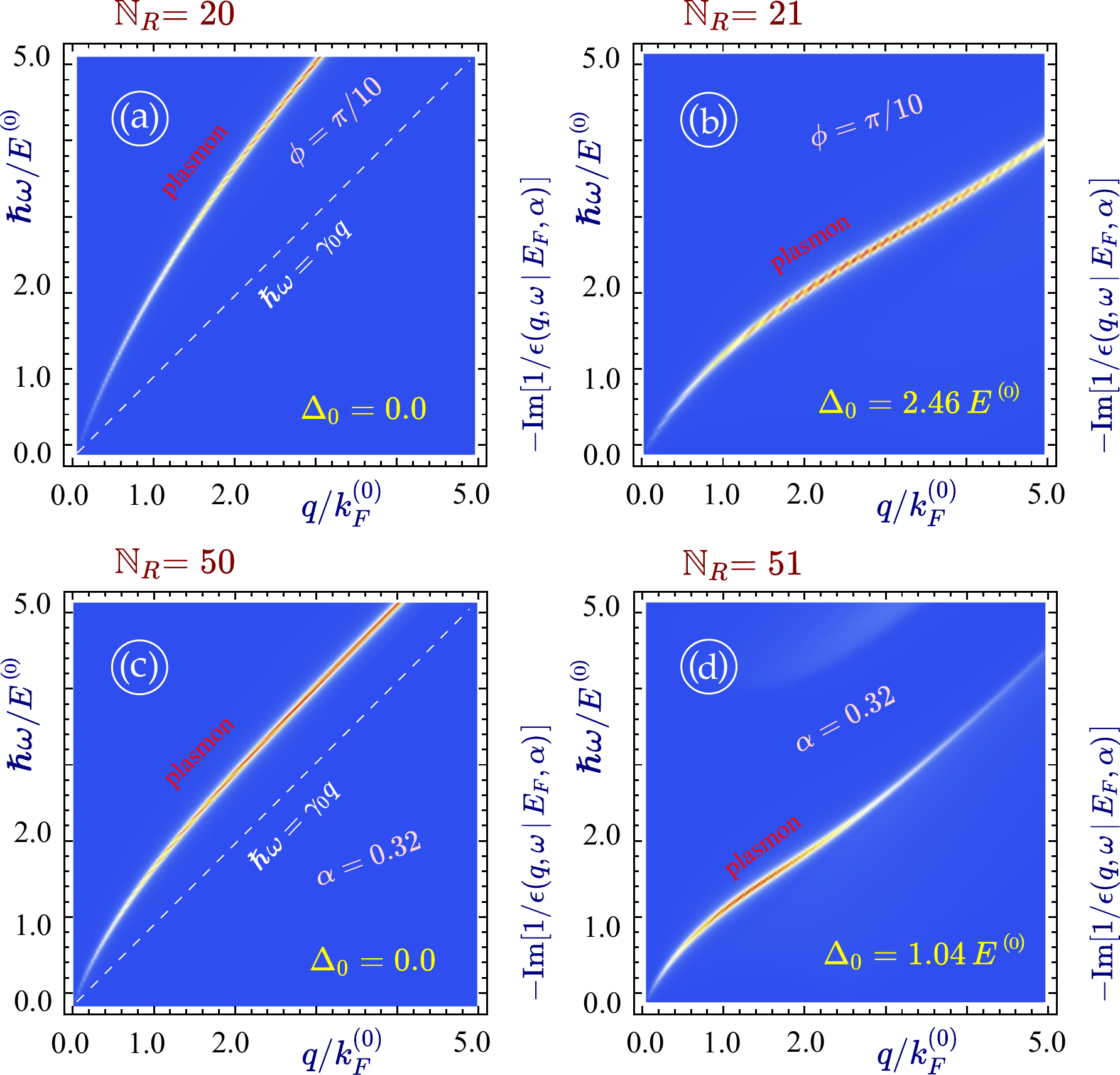}
\caption{(Color online) Plasmon dispersions for nanoribbons with different number $\mbb{N}_R$ of atomic rows across. The density plots of the spectral loss function $\mbb{S}(q, \omega \, \vert \, E_F, \alpha) =  - \text{Im}\left[1/\epsilon(q, \omega \, \vert \, E_F, \alpha)\right]$
are shown for $(a)$ $\mbb{N}_R = 20$, $(b)$ $\mbb{N}_R = 21$, $(c)$ $\mbb{N}_R = 50$ and $(d)$ $\mbb{N}_R = 51$ as a function of frequency $\omega$ and wave vector $q$. Parameter $\alpha$ was chosen $0.32$ (phase $\phi=\pi/10$) for all plots.
}
\label{FIG:5}
\end{figure}

\begin{figure}
 \centering
 \includegraphics[width=0.6\linewidth]{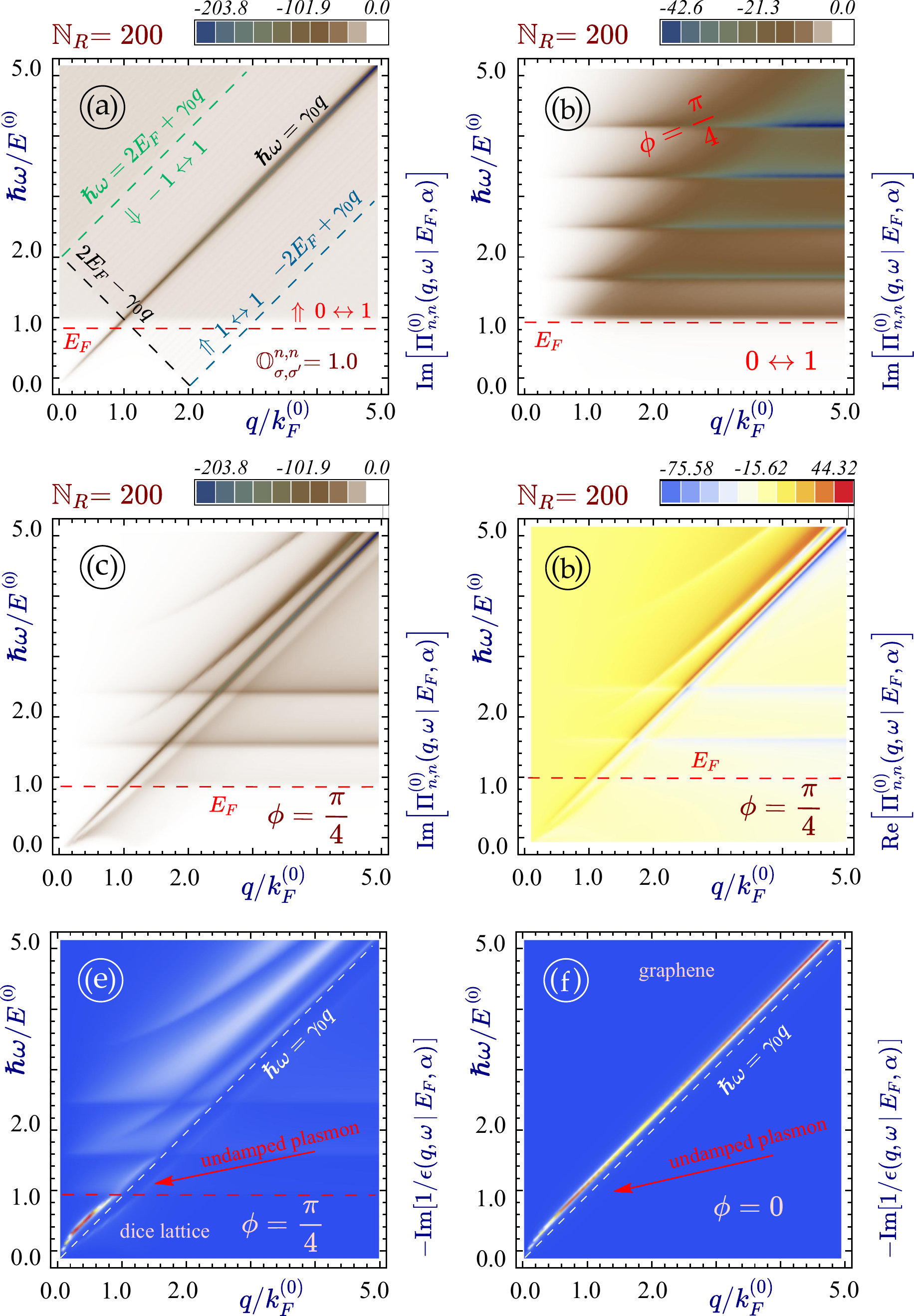}
\caption{(Color online)
Polarization function $\Pi^{(0)}_{n,n'}(q, \omega \, \vert \, E_F, \alpha)$, single-particle excitation spectrum and plasmon excitation dispersions for a wide metallic nanoribbon with $\mbb{N}_R = 200$ and $N_0 = (\mbb{N}_R + 1)/3 = 67$. Panel $(a)$ shows a (unrealistic) 
situation when all the overlaps are equal to unity $\mbb{O}^{\,n,n'}_{\sigma_1 \leftrightarrow \sigma_2} (k_y,q)$ ($\sigma_{1,2} = -1,0,+1$) and equally contribute to the polarization function $\Pi^{(0)}(q, \omega \, \vert \, E_F, \alpha)$. Plot $(b)$ shows the contribution by the transitions from and to the flat band \textit{only} by setting $\mbb{O}^{n,n}_{\,\pm 1 \leftrightarrow \pm 1} (k_y,q) = 0$ for a dice lattice with $\phi = \pi/4$. Panels $(c)$ and $(d)$ display the imaginary and real part of $\Pi^{(0)}(q, \omega \, \vert \, E_F, \alpha)$  for a dice lattice when all the nine possible electron transitions ($\pm 1 \leftrightarrow \pm 1$, $0 \leftrightarrow \pm 1$, $\pm 1 \leftrightarrow 0$
and $0 \leftrightarrow 0$) are taken into account. Plots $(e)$ and $(f)$ demonstrate the plasmon dispersions as the peaks of the spectral loss function $\mbb{S}(q, \omega \, \vert \, E_F, \alpha)$ for graphene with $\phi=0$ and a dice lattice with $\phi=\pi/4$, correspondingly. 
}
\label{FIG:6}
\end{figure}
 
\begin{figure}
 \centering
 \includegraphics[width=0.6\linewidth]{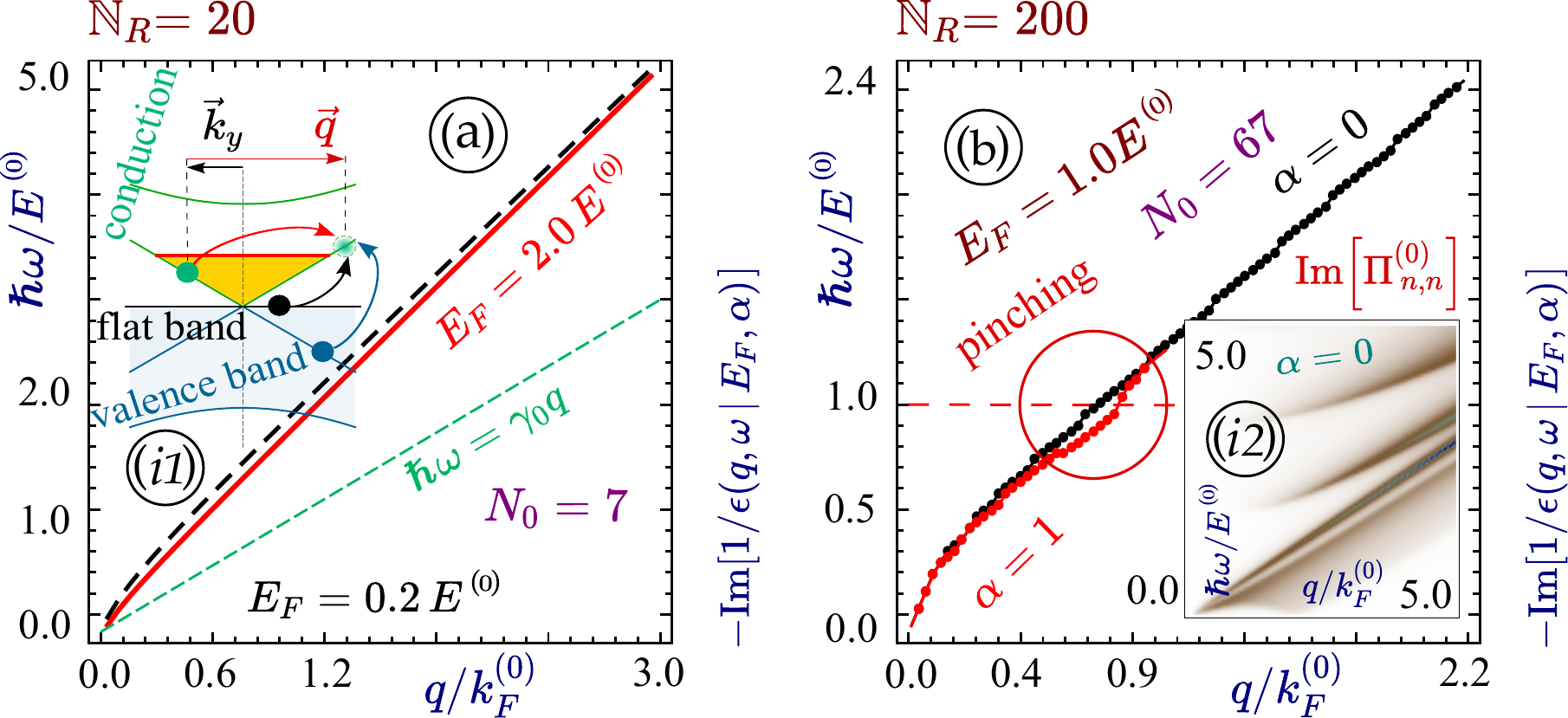}
\caption{(Color online)  Precise adjustments (or fine-tuning) of the plasmon dispersions in $\alpha-\mc{T}_3$-based nanoribbons with armchair edges. Panel $(a)$ demonstrates the plasmons for different values of the electron doping densities $n= ...$ and $...$ which correspond to the Fermi energies $0.2\,E^{(0)}$ and $2.0\,E^{(0)}$. $(b)$ shows the plasmon dispersions for graphene ($\alpha = 0$) and a dice lattice ($\alpha = 1$) for a wide nanoribbon with $\mbb{N}_R = 200$ atomic rows and $E_F =1.0\,E^{(0)} = 93.15\,meV$. Inset $(i1)$ shows the schematic representation of all the allowable electron transitions in a narrow ($\mbb{N}_R = 20$) metallic electron-doped $\alpha-\mc{T}_3$ ribbon, and $(i2)$ describes the single particle excitation spectrum for a wide metallic graphene nanoribbon with $\mbb{N}_R = 20$ and $\alpha = 0$. 
}
\label{FIG:7}
\end{figure}

For an ideal graphene nanoribbon without edge defects, by using the standard many-body theory, the dielectric-function tensor within the random-phase approximation (RPA) can be generally written as

\begin{equation}
\epsilon^{\lambda, \,\rho}_{\mu, \nu}(q, \omega \, \vert \, E_F, \alpha) = 
\delta_{\lambda, \mu}\,\delta_{\rho, \nu} - 
V_{\mu, \nu}^{\lambda, \,\rho}(q) \,\Pi^{(0)}_{\mu, \nu}(q, \omega \, \vert \, E_F, \alpha)\ ,
\label{eqn1}
\end{equation}
where $q$ is the longitudinal transition wave number along a nanoribbon and $\omega$ is the angular frequency of a testing field.  
Each of $\lambda$, $\rho$, $\mu$ and $\nu$ is a composite index which includes subband number and the conduction, valence or flat band 
index, such as $\lambda = \{i,\sigma_i\}$ etc.

\medskip

The plasmon modes of the system can be computed from the determinant of dielectric-function tensor in Eq.\,\eqref{eqn1}, i.e.,

\begin{equation}
 {\cal D}et \, \left[ \epsilon^{\lambda, \,\rho}_{\mu, \nu}(q, \omega \, \vert \, E_F, \alpha) \right] =
 {\cal D}et \, \left[
\delta_{\lambda, \mu}\,\delta_{\rho, \nu} - 
V_{\mu, \nu}^{\lambda, \,\rho}(q) \,\Pi^{(0)}_{\mu, \nu}(q, \omega \, \vert \, E_F, \alpha)
\right] =
 0 \, 
\label{eqn12}
\end{equation}
which is in general could be related to a multi-dimensional matrix $\epsilon^{\lambda, \,\rho}_{\mu, \nu}(q, \omega \, \vert \, E_F, \alpha)$. The diagonal matrix elements in Eq.\,\eqref{eqn1} are expected to give rise to the dispersion of individual plasmon modes, while the off-diagonal matrix elements in Eq.\,\eqref{eqn1} describe the couplings between different plasmon modes.

\medskip
The subband polarization function $\Pi^{(0)}_{\mu, \nu}(q,\omega)$, defined as

\begin{equation}
\Pi^{(0)}_{\mu, \nu}(q,\omega \, \vert \, E_F, \alpha)=\frac{g_s}{2\pi}\,\int\limits_{1^{st}\,{\rm BZ}} dk_y\,
\left\{\frac{
f_0[\varepsilon_\mu(k_y)]-f_0[\varepsilon_\nu(k_y+q)]
}{
\hbar(\omega+i\delta) + \varepsilon_\mu(k_y) - \varepsilon_\nu(k_y+q)
}
\right\}
\,\mbb{O}_{\sigma_\mu \leftrightarrow \sigma_\nu}^{m,n}(k_y, q \, \vert \,  \alpha )\ ,
\label{eqn2}
\end{equation}
where the integral with respect to wave number $k_x$ is limited to the first Brillouin zone, $g_s=2$ takes into account the spin degeneracy, $\delta\ll\omega$ is associated with a homogeneous diagonal-dephasing rate of electrons
$\varepsilon_\mu(k_y)$ is the subband energy, $f_0(x)=\{1+\exp[(x-u_0)/k_BT)]\}^{-1}$ is the Fermi function for thermal-equilibrium electrons, $T$ is the temperature, and $u_0(T)$ is the chemical potential of electrons equal to the Fermi energy $E_F$ at $T=0$ so that 

\begin{equation}
f_0[\varepsilon_\mu(k_y) \, \vert \, E_F, T \rightarrow 0]= \left\{
1+\tet{exp} \left[\frac{\varepsilon_\mu(k_y)-E_F}{k_B T}\right]
\right\}^{-1} \rightarrow \delta_{\sigma_\mu,-1} + \delta_{\sigma_\mu, 1}
\,\Theta[E_F - \varepsilon_\mu(k_y)] \, , 
\end{equation}
where $\Theta(x)$ is a Heaviside step function.  

\medskip 
A crucial simplification comes from the fact that if the mirror symmetry between conduction and valence bands is maintained, as for ideal graphene nanoribbons, the orbital part of wave function becomes independent of band index. Consequently, the Coulomb interaction in 
Eq.\,\eqref{eqn1} is simplified to

\begin{equation}
V^{j, \, m}_{j',\,m'}(q_x)=\frac{e^2}{2\pi\epsilon_0 \, \epsilon_b}\,\int\limits_0^1 du\,\int\limits_0^1 du'\,\cos[\pi(j-m)u]\,\cos[\pi(j'-m')u']\,K_0(|q_x|W|u-u'|)\ 
\label{eqn9}
\end{equation}

is non-negligible only for  $j-j' = 0$ and $m-m' = 0$, as demonstrated in Fig.~\ref{FIG:3} $(a)$. The relevant matrix elements also do not depend on the band index $\sigma$ or phase $\phi$ and, therefore, must be essentially the same for all kinds of $\alpha-\mc{T}_3$ materials including graphene.\,\cite{f1} In the long-wave limit $q \rightarrow 0$, the Bessel function of the second kind $K_0(|q_x|W|u-u'|)$ 
diverges as $-\log(x)$ which looks qualitatively similar to $\backsimeq 1/q$ in two dimensions. The Coulomb matrix element combines two initial and final states  with the same valley and sublattice indices.

\medskip 
Our initial equation \eqref{eqn1} for the dielectric function is now reduced to 

\begin{equation}
\label{eqn100}
\epsilon^{\lambda = \mu}_{\rho = \nu}(q, \omega \, \vert \, E_F, \alpha) = 
\delta_{\sigma_\mu, \sigma_\nu} \delta_{m,n} - 
V_{0}^{0}(q) \,\Pi^{(0)}_{\nu, \nu}(q, \omega \, \vert \, E_F, \alpha)\ ,
\end{equation}

which is a regular two-dimensional matrix ($3N \times 3N$) where $N_{max}$ is the number of subbands which we take into consideration. 
Only intra-subband $(\mu=\nu)$ transition contribute into the polarization function $\Pi^{(0)}_{\mu_2, \mu_2}(q,\omega)$

\medskip
The obtained matrix in \eqref{eqn100} could be even more simplified because the elements depend only on the row index, i.e, all the off-diagonal elements in the same row are identical. $\Pi^{(0)}_{\nu}(q,\omega)$

\par 
The determinant of a matrix with such composition  

\begin{eqnarray}
\label{f2}
&&  {\cal D}et \, \left[  \delta_{\mu, \nu} - 
V_{0}^{0}(q) \,\Pi^{(0)}_{\, \nu, \nu}(q, \omega \, \vert \, E_F, \alpha)  
\right] = \\
\nonumber 
&& \sum _{\mu_{1},\mu_{2}, \ldots , \,\mu_{3 N_{max}}=1}^{3 N_{max}}
\mc{L}_\varepsilon (\mu_{1},\mu_{2}, \ldots , \,3 N_{max}) \,
\epsilon^{\mu}_{\nu}(q, \omega \, \vert \, E_F, \alpha)_{1,\mu_1} \cdots \epsilon^{\mu}_{\nu}(q, \omega \, \vert \, E_F, \alpha)_{3 N_{max},\,
\mu_{3 N_{max}}} \, , 
\end{eqnarray}

where $\mc{L}_\varepsilon (\mu_{1},\mu_{2}, \ldots , \,3 N_{max})$ is the Levi-Civita tensor, is simplified to a single summation over the only remaining composite index $\nu = \{ \sigma_\nu, n\}$ 

\begin{equation}
\label{fdet}
 {\cal D}et \, \left[ \epsilon^{\mu}_{\nu}(q, \omega \, \vert \, E_F, \alpha) \right] = 1 - V_{0}^{0}(q) \, \, 
\sum\limits_{n=1}^{N_{max}} \Pi^{(0)}_{n,n}(q, \omega \, \vert \, E_F, \alpha) \, , 
\end{equation}

where $\Pi^{(0)}_{n,n}(q,\omega)$ is the subband polarizability which 
already includes the summation over the valence, conductance and flat bands

\begin{equation}
\label{pin}
\Pi^{(0)}_{n,n}(q, \omega \, \vert \, E_F, \alpha) =
\sum\limits_{\sigma={\pm 1, 0}} \Pi^{(0)}_{\, \nu,\nu}(q, \omega \, \vert \, E_F, \alpha) \, . 
\end{equation}

This is especially straightforward to verify for the case of a smallest possible $2 \times 2$ matrix

\begin{eqnarray}
\label{f3}
&&  {\cal D}et \, \left[ 
\begin{array}{cc}
1 - V_{0}^{0}(q) \, \Pi^{(0)}_{\, 1, 1}(q, \omega \, \vert \, E_F, \alpha)  & - V_{0}^{0}(q) \,
\Pi^{(0)}_{\, 2, 2}(q, \omega \, \vert \, E_F, \alpha) \\
- V_{0}^{0}(q) \, \Pi^{(0)}_{\, 1, 1}(q, \omega \, \vert \, E_F, \alpha) & 1 - V_{0}^{0}(q) \, 
\Pi^{(0)}_{\, 2, 2}(q, \omega \, \vert \, E_F, \alpha)
\end{array}
\right] = \\
\nonumber 
&& = \left[1 - V_{0}^{0}(q) \Pi^{(0)}_{\, 1, 1}\right] \, \left[1 - V_{0}^{0}(q) \Pi^{(0)}_{\, 2, 2}\right] - \left[V_{0}^{0}(q)\right]^2 
\, \Pi^{(0)}_{\, 1, 1} \Pi^{(0)}_{\, 2, 2} = 1 - V_{0}^{0}(q) \, \left[ \Pi^{(0)}_{\, 1, 1} + \Pi^{(0)}_{\, 2, 2}\right] \, . 
\end{eqnarray}

Relation \eqref{fdet} for a general $N \times N$ could be proven using the method of induction. We begin from the Laplace expansion
of matrix \eqref{eqn100} over its last row  $\left\{ - V_{0}^{0}(q) \, \Pi^{(0)}_{\, 3 N_{max}, 3 N_{max}}, \, - V_{0}^{0}(q) \, \Pi^{(0)}_{\, 3 N_{max}, 3 N_{max}} \, \cdots \, 1 - V_{0}^{0}(q) \, \Pi^{(0)}_{\, 3 N_{max}, 3 N_{max}} \right\}$:

\begin{eqnarray}
\label{A}
&& {\cal D}et \, 
\left[ \epsilon^{\mu}_{\nu}(q, \omega \, \vert \, E_F, \alpha) \right]
=\sum _{\mu=1}^{3 N_{max}} (-1)^{\mu + 3 N_{max}} \epsilon^{\mu}_{3 N_{max}} \, \mc{M}_{\mu, 3 N_{max}} = 
\\ 
\nonumber 
&& 
V_{0}^{0}(q) \,  \sum _{\mu=1}^{3 N_{max} - 1 } (-1)^{\mu + 3 N_{max}} V_{0}^{0}(q) \, \Pi^{(0)}_{\, \mu, \mu} \, \mc{M}_{\mu, 3 N_{max}} 
+ \left[ 1- V_{0}^{0}(q) \, \Pi^{(0)}_{\, 3 N_{max}, 3 N_{max}} \right] \, \mc{M}_{3 N_{max}, 3 N_{max}} 
  \, , 
\end{eqnarray}

where the corresponding minor matrices are calculated as 

\begin{eqnarray}
\label{w1}
&& \mc{M}_{\mu, 3 N_{max}} = - V_{0}^{0}(q) \, \Pi^{(0)}_{\mu, 3 N_{max}}(q, \omega \, \vert \, E_F, \alpha) 
\hskip0.2in \text{for} \hskip0.2in \mu \neq 3 N_{max} 
\\
\nonumber
&& \text{and} 
\\
\label{w2}
&& \mc{M}_{3 N_{max} , 3 N_{max}} = 1 -  V_{0}^{0}(q) \, \sum\limits_{\mu = 1}^{3 N_{max} -1}
 \Pi^{(0)}_{\mu, 3 N_{max}}(q, \omega \, \vert \, E_F, \alpha) \, . 
\end{eqnarray} 

The validity of Eq.~\eqref{w2} is assumed as the base of the induction. The complete summation in Eq.~\eqref{A} amounts to 
expression \eqref{fdet}. However, we must say that a complete proof of \eqref{fdet} falls off the scope of the present paper. 
Finally, the determinant $\mc{D}et$ and the trace $\mc{T}r$ of matrix \eqref{fdet} are connected as 

\begin{equation}
\mc{D}et \left[ \epsilon^{\mu}_{\nu}(q, \omega \, \vert \, E_F, \alpha) \right] = 
\mc{T}r \left[ \epsilon^{\mu}_{\nu}(q, \omega \, \vert \, E_F, \alpha) \right] - 3 N_{max} + 1 \, ,
\end{equation} 

which was also employed in Ref.~[\onlinecite{f1}] for graphene. In our calculation, the actual polarization function is obtained as

\begin{equation}
\Pi^{(0)}(q, \omega \, \vert \, E_F, \alpha) = \sum\limits_{n = N_0-10}^{N_0+10} \Pi^{(0)}_{n,n'}(q, \omega \, \vert \, E_F, \alpha)
\end{equation}
around the lowest subband $N_0$.

\medskip
Our next step is to calculate the wave function overlaps (prefactors) $\mbb{O}^{\,n,n'}_{\pm 1,0 \leftrightarrow \pm 1,0} (k_y,q)$ which 
enter Eq.~\eqref{eqn2} for the polarization function. For a $\alpha-\mc{T}_3$ nanoribbon, it is calculated as

\begin{equation}
\mbb{O}^{\,n,n'}_{\pm 1,0 \leftrightarrow \pm 1,0} (k_y, q) 
\equiv\left| \langle  \Phi^\phi_{\gamma}(n \, \vert \, k_y) 
\, \vert \texttt{e}^{i q_x x} \vert \,
\Phi^\phi_{\gamma}(n' \, \vert \, k_y + q) 
\rangle\right|^2
\label{eqn3}
\end{equation}

where the complete wave functions $\Phi^\phi_{\gamma}(n \, \vert \, k_y)$  and $\Phi^\phi_{\gamma}(n' \, \vert \, k_y + q)$ 
given by Eq.~\eqref{wave0}.

\medskip

The wave function overlaps $\mbb{O}^{\,n,n'}_{\pm 1,0 \leftrightarrow \pm 1,0} (k_y, q)$ obtained from Eq.~\eqref{eqn3}  

\begin{eqnarray}
\label{p11}
&& \mbb{O}^{\,n,n'}_{\pm 1 \leftrightarrow 1} (k_y,q) = \frac{1}{4}\, \left[
1 \pm \cos [\Theta_{n,n'}(k_y,q)]
\right]^2 + \frac{1}{4} \, \cos^2 (2 \phi) \sin^2 [\Theta_{n,n'}(k_y,q)]
\end{eqnarray}
and 

\begin{eqnarray}
\label{p01}
&& \mbb{O}^{\,n,n'}_{0 \leftrightarrow 1} (k_y,q) = \frac{1}{2}\, \sin^2 (2 \phi) \, \sin^2 [\Theta_{n,n'}(k_y,q)]
\end{eqnarray}
are the same as we earlier obtained for the bulk $\alpha-\mc{T}_3$ \,\cite{ourplay} since $\mbb{O}^{\,n,n'}_{\pm 1,0 \leftrightarrow \pm 1,0} (k_y, q)$ do not depend on the valley index $\tau$ and for an armchair nanoribbon wave function \eqref{wave0} is just a combination of the states from the valleys. Therefore, an overlap \eqref{eqn3} is essentially an average between the two inequivalent valleys. At the same time, 
form factors \eqref{wave0} directly depend on the band indices so that the obtained result \eqref{p11} and \eqref{p01} are completely different 
for the various type of carrier transitions (from and to valence, flat and conduction bands).

\medskip
For graphene with $\alpha = 1$, we easily recover $\mbb{O}^{\,n,n'}_{\pm 1 \leftrightarrow \pm 1} (k_y,q) = 1/2 (1 \pm \cos [\Theta_{n,n'}(k_y,q)])$ obtained in Ref.~[\onlinecite{f2}].  For the opposite limiting case of a dice lattice, we obtain $1/4 (1 \pm \cos [\Theta_{n,n'}(k_y,q)])^2$. 

\medskip
We should emphasize that overlaps \eqref{p11} and \eqref{p01} are mainly responsible for all the crucial difference between the plasmons in nanoribbons and bulk $\alpha-\mc{T}_3$ materials. Angle $\Theta_{n,n'}(k_y,q)$ between the wave vectors $k_y$ and $k_y + q$ is quantized due to the quantization of allowable values of the transverse momenta $\xi_{n\,(n')}$. The longitudinal momenta $k_y$ and $k_y + q$ are always directed along the $y-$axis only. Since the biggest contribution to the polarization function comes for several lowest subbands with the minimal values of $\xi_{n\,(n')}$, the calculated angles $\Theta_{n,n'}(k_y,q)$ are often close to $0$ or $\pi$ so that large number of relevant overlaps are equal to either 1 or 0, especially in the long wave limit with a small transfer momentum $q$. This is drastically different from the corresponding bulk material.

\section{Results and discussion}
\label{s04}

We begin our discussion with the case of an armchair $\alpha-\mc{T}_3$ nanoribbon with metallic (gapless) subbands in which 
in which the lowest and and the next consequent subbands are far removed.\,\cite{f2, mr1, mr2} In general, the situation with 
the energy separation for the quantized subbands in a finite-width ribbon is qualitatively similar to a quantum well, i.e., 
such well-separated levels are found for a narrow ribbon.

\par 
The transverse momenta for the electrons at the Fermi surface are negligible and the angle $\theta_n$ associated with wave vector $\{\xi_n, k_y\}$ are  

\medskip

\begin{equation}
\theta_n(k_y) = \left\{
\begin{array}{c}
-\pi/2    \hskip0.1in \text{if}  \hskip0.1in k_y < 0 \\
\hskip0.17in \pi/2 \hskip0.1in \text{if} \hskip0.1in k_y > 0 \, ,
\end{array}
\right.
\end{equation}

and for angle  $\Theta_{n,n'} (k_y,q)$ between the two different states $\{\xi_n, k_y\}$ and $\{\xi_n, k_y + q\}$ which enters the prefactors \eqref{p11} and \eqref{p01} 

\begin{equation}
\Theta_{n,n'} (k_y,q) = 
\left\{
\begin{array}{l}
-\pi \hskip0.1in \text{if}  \hskip0.1in k_y > 0 \,\,\&\,\, k_y + q < 0 \, , \\
\hskip0.15in 0 \hskip0.1in \text{if}  \hskip0.1in k_y (k_y + q) > 0 \, , \\
\hskip0.15in \pi \hskip0.1in \text{if}  \hskip0.1in k_y < 0 \,\,\&\,\, k_y + q > 0 \, . \\
\end{array}
\right.
\end{equation}

In all these cases, $\sin [\Theta_{n,n'}(k_y,q)] = 0$, $\mbb{O}^{\,n,n'}_{0 \leftrightarrow 1} (k_y,q) = 0$ and 
$\mbb{O}^{\,n,n'}_{\pm 1 \leftrightarrow 1} (k_y,q) = 1/4 (1 \pm \cos [\Theta_{n,n'}(k_y,q)])^2$ is equal to either 0 
or 1 according to

\begin{equation}
\mbb{O}^{\,n,n'}_{\sigma \leftrightarrow 1} (k_y,q)= 
\left\{
\begin{array}{l}
0 \hskip0.1in \text{if}  \hskip0.1in \sigma k_y (k_y + q) < 0 \, , \\
1 \hskip0.1in \text{if}  \hskip0.1in \sigma k_y (k_y + q) > 0 \, . \\
\end{array}
\right.
\end{equation}

 The obtained polarization function does not depend on $\phi$, and the whole situation is exactly
similar to what was earlier found for graphene.\,\cite{mr1,f1}

\par 

Therefore, the plasmons in the long-wave limit for $\alpha-\mc{T}_3$ are given by \,\cite{mr1}
\begin{equation}
\Omega_p(q) = \sqrt{ - q \, \left(q + \frac{2 E_F}{\gamma_0} \right) \, \ln (q W_R)} \, . 
\end{equation}
The obtained result implies that for the case described above the plasmon dispersions in the long-wave limit do not 
depend on $\alpha$ and are exactly the same for all $\alpha-\mc{T}_3$ lattices, including the limiting cases of a dice lattice
and graphene. 

\par 
\medskip
\medskip
\par 

For all the other cases, wave function overlaps \eqref{p11} and \eqref{p01} are finite and depend on $q$. This occurs if angle $\Theta_{n,n}(k_y,q)$ different from 0 and $\pi$. This is achieved for a finite $\xi_n$ disregarding of the ribbon width. For a wide ribbon, the situation is close to a bulk $\alpha-\mc{T}_3$ since the increment in $\xi_n$ between the subband levels is decreased. The effect of the transfer wave vector $q$ which is normally considered limited to $\backsim k_F$ or much less than that in the long wave limit also depends on the magnitude of initial $k_y$ value and its relation to $\xi_n$, as we can see in Fig.~\ref{FIG:3}.    

\par 
The difference between graphene and other types of $\alpha-\mc{T}_3$ lattices obviously originates from the transitions associated with the flat band. \,\cite{malpd} Since the form factors \eqref{p01} for such transitions are $\backsim \sin^2 (2 \phi) \, \sin^2 [\Theta_{n,n}(k_y,q)]$, they are severely suppressed for all $\alpha \backsimeq 0$. i.e., for all materials close to graphene and for small transverse components 
$\xi_n$ which is always the case for a narrow ribbon. Therefore, we conclude that the effect of parameter $\alpha$ in our considered range of $q$ is noticeable only for a sufficiently wide ribbon. If $q$ is set large enough $\backsim 10 k_F$ (see Fig.~\ref{FIG:3} $(d)$) overlaps \eqref{p01} become observable for any width of the ribbon.

\medskip
The real and imaginary parts of the dynamical polarization function for armchair nanoribbons are presented in Fig.~\ref{FIG:4} and the 
corresponding plasmon dispersions - in Fig.~\ref{FIG:5}. We see that energy bandgap plays a crucial role in shaping the plasmon and its
damping for all widths of the nanoribbon. The single-particle excitation region is split into several areas corresponding to the different 
subbands in addition to the main split separating $-1 \leftrightarrow 1$ and 2DEG with two parabolic edges.

\par 
The plasmon frequencies are decreased in the presence of the gap for all $q$. This effect has some similarity to $\backsimeq \left( 1 - \Delta_0^2/E_F^2 \right)$ dependence in the bulk, but for a nanoribbon this reduction is less. For larger values of $q$, the plasmon branch tends to the lower, intra-band section of the particle-hole modes and a special concave-convex shape of the plasmon dispersion curve is found (see Fig.~\ref{FIG:5} $(b)$ and $(d)$).

\medskip
For larger $\mbb{N}_R$, the upper inter-band single particle excitation region (Inter-SPE) is observed within our range of the wave vector $q$ and frequency $\omega$. In the case of zero gap, the single-particle excitation regions are reduced into a small area along the main diagonal $\omega = \gamma_0 q$ and the plasmon is not damped.

\par 
Fig.~\ref{FIG:6} shows the effect of relative hopping parameter $\alpha$ for a wide nanoribbon with $\mbb{N}_R = 200$. The energy separation between the two closest subbands reduces to $\backsim 0.7 E^{(0)}$  which comparable with the regular doping density. As a result, several subbands become populated and the situation resembles a bulk material. 

\par 
First, we examine a (unrealistic) situation when all the form factors are equal to unity. We clearly see the contributions from all possible types of  transitions and a strong peak around the main diagonal $\hbar \omega = \gamma_0 q$.  When the flat band contribution become important, a plasmons encounters additional Landau damping due to a particle-hole mode located at $\hbar \omega \geq E_F$ which appears due to the electron transition between the conduction and flat bands. The contribution from such transitions shown in Fig.~\ref{FIG:6} $(b)$ is not uniform but is built of several discrete pieces due to each subband level. Its contribution to $\text{Im} \left[\Pi^{(0)}_{n,n}(q, \omega) \right]$ at the Fermi level is about 1/2 of its maximum value. It obviously increases at larger frequencies which is expected since more levels become relevant. Also, it is larger for larger values of wave vector $q$ for the same frequency. Such a strong plasmon damping at relatively low $q$ constitutes the most drastic difference between graphene and $\alpha-\mc{T}_3$ for both bulk and the nanoribbons (see 
(Figs.~\ref{FIG:6} $(e)$ and $(f)$). The plasmon branch also demonstrates a special shape with a pinching to a single point $\Omega_p = q = E_F$. The strength of this pinching is increased for a larger $\alpha$ (see Fig.~\ref{FIG:7} $(b)$).

\medskip 
Importantly, we see that the plasmon frequency is much less sensitive to the electron doping (doping density or the chemical potential) than in the case of bulk graphene or a dice lattice, as shown at Fig.~\ref{FIG:7} $(a)$. Indeed, the most crucial $E_F$-dependent term of the polarization function \eqref{eqn2} $1 \leftrightarrow 1$ which monotonically depends on $E_F$ and would disappear if $E_F = 0$ in fact becomes independent of $E_F$ if $E_F < q$ since this is always a transition from an occupied to a free state or zero. Also, there is no averaging by the direction which we needed to perform in bulk $\alpha-\mc{T}_3$.

\section{Summary and Remarks} 
\label{s05}

In this paper, we have calculated the dynamical polarization function, plasmon dispersions and Landau damping for the various types of $\alpha-\mc{T}_3$-based armchair nanoribbons. Specifically, we have considered the principally different cases of zero and finite energy gap between the valence and conduction bands which is regulated by the width or the number of atomic rows across the ribbon. 

\par
In the case of a finite gap, the inter- and intra-band portions of the particle-hole modes are split into two separate regions similarly to the bulk case. However, for a finite-width ribbon each mode is also divided into the distinguished discrete areas which correspond to the quantized energy levels of the single-electron dispersions. The location and intensity of these separate regions depends on the width of the ribbon. In particular, they could be very far removed for a narrow AGNR. The plasmon branch is located at the lower frequencies for the case of a finite gap but this effect is less noticeable compared to bulk graphene or a dice lattice. 

\medskip 
The most substantial difference between the bulk $\alpha-\mc{T}_3$ and a ribbon stems from the wave function overlaps (prefactors) which depend on a single subband index $n$ and most importantly become equal or close to zero for a number of relevant transitions in the case of a small ribbon width. Thus, we see that for $\mbb{N}_R \leq 20$ the obtained plasmon dispersion show almost no dependence on relative hopping parameter $\alpha$, i.e., the situation is nearly the same for graphene and all types of $\alpha-\mc{T}_3$ materials including a dice lattice.  However, more energy levels need to by taken into account in a wider ribbon since for a finite-level doping several subband could become populated. Apart from that, one cannot rely on the Dirac Hamiltonian approximation for a very narrow ribbon. The previously considered situation with a small doping and a large energy separation between the lowest metallic and its nearest neighbor subband is important but limited and cannot be applied for most realistic problems which motivated our present study. 

\par 

We have also considered a wide ribbon with $\mbb{N}_R = 200$ in which the $0 \leftrightarrow 1$ transition associated with the flat band become substantial, and found a strong dependence of the obtained plasmons on $\alpha$. First, the plasmon branch is damped at the Fermi level because of an additional piece of the particle hole modes. We also see a pinching of the plasmon dispersions around the 
Fermi level. Similar phenomena have been earlier observed in the bulk $\alpha-\mc{T}_3$ but in a ribbon these effects could be regulated from negligible to a very strong level by adjusting the ribbon width and the bandgap.

\medskip
In conclusion, we have performed a comprehensive study of the plasmon excitations in $\alpha-\mc{T}_3$ nanoribbons and uncovered some very special types of dependence of the plasmon frequencies and the regions of finite Landau damping on the width of the ribbon, absence or presence of the energy bandgap, electron doping density and parameter $\alpha$ of an $\alpha-\mc{T}_3$ lattice. We are confident that these novel and earlier unseen plasmon dispersions will be widely employed in creating a new generation of nanoribbon-based electronic, optical and plasmonic devices.

\section{Acknowledgements}
A.I. would like to acknowledge the funding received from TRADA-51-82 PSC-CUNY Award $\#$ 63061-00-51. D.H. was supported by the Air Force Office of Scientific Research (AFOSR).  G.G. would like to acknowledge Grant No. FA9453-21-1-0046 from the Air Force Research Laboratory (AFRL).

\bibliography{RibBib} 

\end{document}